# Understanding Charge Transport in Single Molecule of Rhenium(I) Compounds: A Computational Approach

Rajwinder Kaur[1], Savas Kaya[2], Konstantin P. Katin[3,][*] and Prakash Chandra Mondal[1,][*]

[1]Department of Chemistry, Indian Institute of Technology Kanpur, Uttar Pradesh-208016, India

[2]Department of Pharmacy, Faculty of Science, Cumhuriyet University, Sivas 58140, Turkey

[3]Laboratory of 2D Nanomaterials in Electronics, Photonics and Spintronics, National Research Nuclear University "MEPhI", 31 Kashirskoe sh., 115409 Moscow, Russia

E-mail: pcmondal@iitk.ac.in (P.C.M.)

**Abstract**

Understanding electrical characteristics and corresponding transport models at single molecular junctions is crucial. There have been many reports on organic compounds-based single molecular junctions. However, organometallic compounds-based single molecular junctions have not been explored yet. Re(I) organometallic compounds are known to exhibit intriguing photophysical properties scrutinized for photocatalysis, and light-emitting diodes but have not been explored in molecular electronics. In this work, a theoretical model study on the I-V characteristics of two Re(I)-carbonyl complexes bearing Re-P and Re-N, N linkage has been meticulously chosen. Tunneling and hopping transport in Au/Re(I)-complex/Au single-molecule junctions are governed by Landauer-formalism and the Marcus theory, respectively. Interestingly, variations in molecular architecture culminate in notable variations in junction functionality and mechanism of charge conduction. Physical parameters influencing the device characteristics such as dipole moment, molecule-electrode coupling strength, voltage division factor, and temperature have been extensively studied which offers modulation of the characteristics and device design. The dominant hopping current in Re complex bearing bipyridine linkage was found to be responsible for the observed asymmetric electrical (I-V) behavior. Our work paves the way for constructing various organometallic compounds-based molecular junctions to understand electronic functions and the underlying transport mechanisms.

**Keywords:** Re(I) carbonyl complex, single-molecule junction, electrical characteristics, charge transport, tunneling,

**Introduction**

Molecular electronics (MEs) is an emerging field, where molecules can serve the purpose of active electronic elements such as wire, diode, switch, transistor, etc.[1–6] Molecules are considered the second smallest objects after atoms. It is a formidable and expensive method to make junctions at the atomic scale. However, molecules offer many advantages in studying charge transport varying structures, compositions, and functionalities thus facilitating miniaturization in size where molecules act as electronic circuit elements, mimicking complementary metal-oxide-semiconductor (CMOS) technology. Advancements in molecular electronics commenced after the first theoretical framework proposed by Aviram and Ratner on molecular rectifiers, back in 1974.[7] Single-molecule junctions are the ultimate goal of molecular electronics for understanding the electronic functions and charge transport models.[8,9] There are enormous challenges in bridging a single molecule between two electrodes to form stable molecular junctions (MJs). The development of scanning tunneling and atomic force microscopy emerged as a boon for analyzing single-molecule junctions.[10–12] The advantage of dealing with molecules includes the accessible anchoring groups such as thiols, silanes, carboxylates, primary amine, and pyridine that can be utilized in making nanoscale molecular layers growth on various substrates via a self-assembly process.[13,14] The self-assembly of molecules on the surface of the desired substrate is contingent upon the molecule-substrate interactions.[15–17] In molecular junctions, the gold (Au)-thiol (-SH) self-assembly systems have been well explored.[18] The Au electrode and thiols contain molecules in which soft-soft interactions form stable electrode-molecule interfaces, thus many researchers consider self-assembled monolayers ideal platforms for fabricating nanoelectronic molecular junctions with a two-terminal device stacking configuration of Au/thiolated molecule/Au. The extent of molecule-substrate interactions decides the formation of hybrid states at the electrode-molecule interface which can modulate the energy landscape of either Fermi energy ($E_F$) of the electrodes or frontier molecular orbitals such as HOMO, and LUMO which influence the charge conduction and subsequently the transport model.[19–23] To realize molecular electronic devices, it is imperative to understand the mechanism of charge transport at the molecular level.[24–28] Tunneling and hopping are well-established charge conduction mechanisms, where tunneling persists in small-size molecular layers with a thickness of $d \sim 5$ nm and is invariable to temperature while hopping conduction is a thermally activated process that exists in molecular films thickness of $d > 5$ nm.[29] Redox-active hopping is well-known in metal complexes-based molecular junctions.[30–32] However, there can be a transition in the mechanism upon the increase in the molecular length that separates two macroscopic electrical contacts. For instance, the Frisbie group demonstrated the transition from tunneling to hopping charge conduction in conjugated oligophenyleneimine wires by measuring resistance as a function of length, temperature, and applied bias.[33] The Tao group has performed length and temperature dependence of conductance curves in single-molecule junctions measured employing the scanning tunneling microscope-based break-junction (STM-BJ) method has shown the higher conductance at elevated temperatures observed in the case of long molecular wires attributed to the hopping mechanism.[34] The crossover point for the transition from tunneling to hopping has been reported at molecular lengths of 5.2-7 nm, 4 nm, 3 nm, and 2.7 nm for dithiolates, oligophenyleneimine thiolates, oligo arylene ethynylene derivatives, and oligo phenylene-ethynylene, respectively.[33–36] One mechanism can dominate over the other in a particular set of conditions, however, both tunneling and hopping contribute to the total current observed in molecular junctions. The molecular junctions utilize metal-complexes such as Fe (II), Ru (II), Os (II), and Co (II)-polypyridyl complexes were experimentally investigated for charge transport in molecular junctions, Ru (II) being the most studied ones.[37,38]

Re (I) (a $d^6$ metal ion) carbonyls-based organometallic compounds are of great interest for many applications due to their photophysical properties. They have been explored in photocatalysis, carbon dioxide reduction, and light-emitting diodes.[39–41] However, such compounds have not been considered for quantum mechanical study in single-molecule junctions. In this work, the theoretical investigation of the charge transport mechanism in two chemically different rhenium (I) carbonyl complexes bridged between gold electrodes via thiol linkage forming a single molecule junction has been conducted. Coherent tunneling has been thoroughly modeled using Landauer formalism to different molecular orbitals. The hopping mechanism has been described based on the Marcus theory, which considers the rate-determining step, reorganization energies, and temperature dependence.[42] As per theoretical results, molecule 1 or M1 ($ReC_{28}PH_{19}S_3O_7$) showcased symmetric current-voltage (I-V) behavior while molecule 2 or M2 ($ReC_{21}N_2H_{12}S_3O_6$) exhibited asymmetric I-V or rectifying properties. The dominance of the hopping mechanism was found to give rise to the rectification character of the molecular junction.[43] Theoretical investigations on such systems could be of great significance in the design of molecules for rectification purposes.

**Computational details**

Geometries of all finite-size structures were optimized in the frame of the density functional theory GPU-based TeraChem program.[44,45] A 6-311G* basic set for organic ligands was combined with the lanl2dz set for metal atoms (Re and Au).[46,47] Common B3LYP functional was applied to all elements.[48,49] Post-processing of the molecular orbitals was done with MultiWfn software [50]. Fermi energy and electronic density of states of the fcc gold crystal were calculated using the Quantum Espresso software.[51,52] The GGA-PBE method was combined with ultrasoft pseudopotentials from the PS library.[53] Plane-wave basis was used with the energy cutoff of 50 and 400 Ry for wavefunctions and charge density, respectively.

## Results and discussion

### Molecular orbitals

Redox chemistry of Re(I)-carbonyl complexes (Re(I) to Re(II) and vice versa) is well-explored owing to photophysical and photochemical applications attracted to explore its characteristics for molecular devices. For theoretical model studies, two Re(I) carbonyl complexes are considered for charge transport in single-molecule, Au/Re-complex/Au junctions.

We considered two molecules, M1 ($ReC_{28}PH_{19}S_3O_7$) and M2 ($ReC_{21}N_2H_{12}S_3O_6$), presented in **Fig. 1** (see **Fig. S1** for chemical structures of the compounds). Both neutral ($M1^0$ and $M2^0$) and positively charged ($M1^+$ and $M2^+$) forms of the molecules are considered. Structures of both forms were optimized without symmetry constraints and optimized parameters are provided in **Tables S1 and S2**. Energies of frontier molecular orbitals, HOMO, and LUMO, energy gap, and dipole moments of all molecules are presented in **Table 1**.

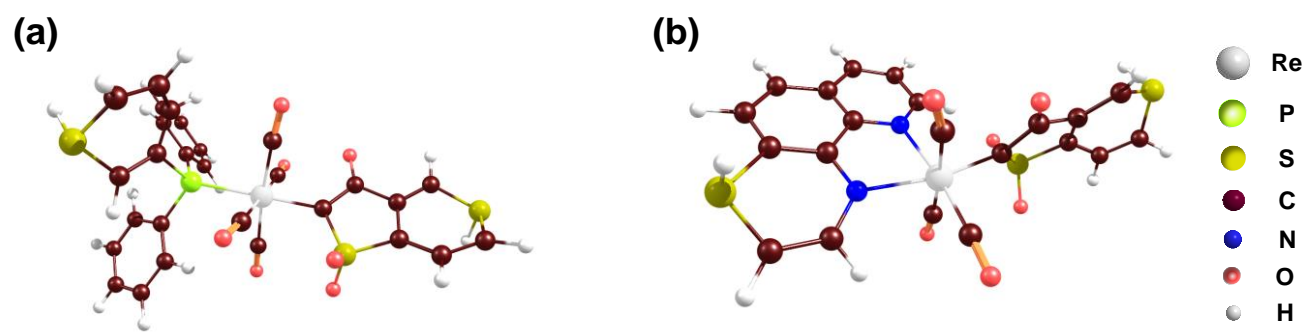

**Fig.1.** Optimized structures of (a) M1, and (b) M2 molecules. The inset shows the color codes of the respective elements of the compounds.

**Table 1**. Energies of HOMO, LUMO, and HOMO-LUMO gap (eV) and dipole moments $D$ (Debye) of the considered molecules.

| Molecule | HOMO (eV) | LUMO (eV) | Gap (eV) | $D$ |
|:---:|:---:|:---:|:---:|:---:|
| $M1^0$ | -3.92 | -2.10 | 1.82 | 5.88 |
| $M1^+$ | -7.37 | -7.13 | 0.24 | 2.94 |
| $M2^0$ | -3.38 | -3.21 | 0.17 | 10.99 |

| M2$^+$ | -7.65 | -6.83 | 0.82 | 5.73 |

Note that not only HOMO and LUMO but other orbitals can contribute to charge transport in response to the applied bias. Therefore, we consider eight molecular orbitals and their energies for charged M1$^+$ and M2$^+$ molecules, from HOMO – 3 to LUMO + 3. Their energies are presented in **Table 2**, whereas their shapes are plotted in **Figs. 2** and **3**.

**Table 2**. The energy of considered molecular orbitals of the charged M1$^+$ and M2$^+$ molecules is determined utilizing Density Functional Theory.

| **Orbital** | **Entry** | **M1$^+$** | **M2$^+$** |
|:---:|:---:|:---:|:---:|
| HOMO – 3 | 1 | -8.65 | -8.91 |
| HOMO – 2 | 2 | -8.38 | -8.48 |
| HOMO – 1 | 3 | -7.73 | -7.93 |
| HOMO | 4 | -7.37 | -7.65 |
| LUMO | 5 | -7.13 | -6.83 |
| LUMO + 1 | 6 | -5.18 | -5.97 |
| LUMO + 2 | 7 | -5.12 | -5.25 |
| LUMO + 3 | 8 | -5.00 | -5.05 |

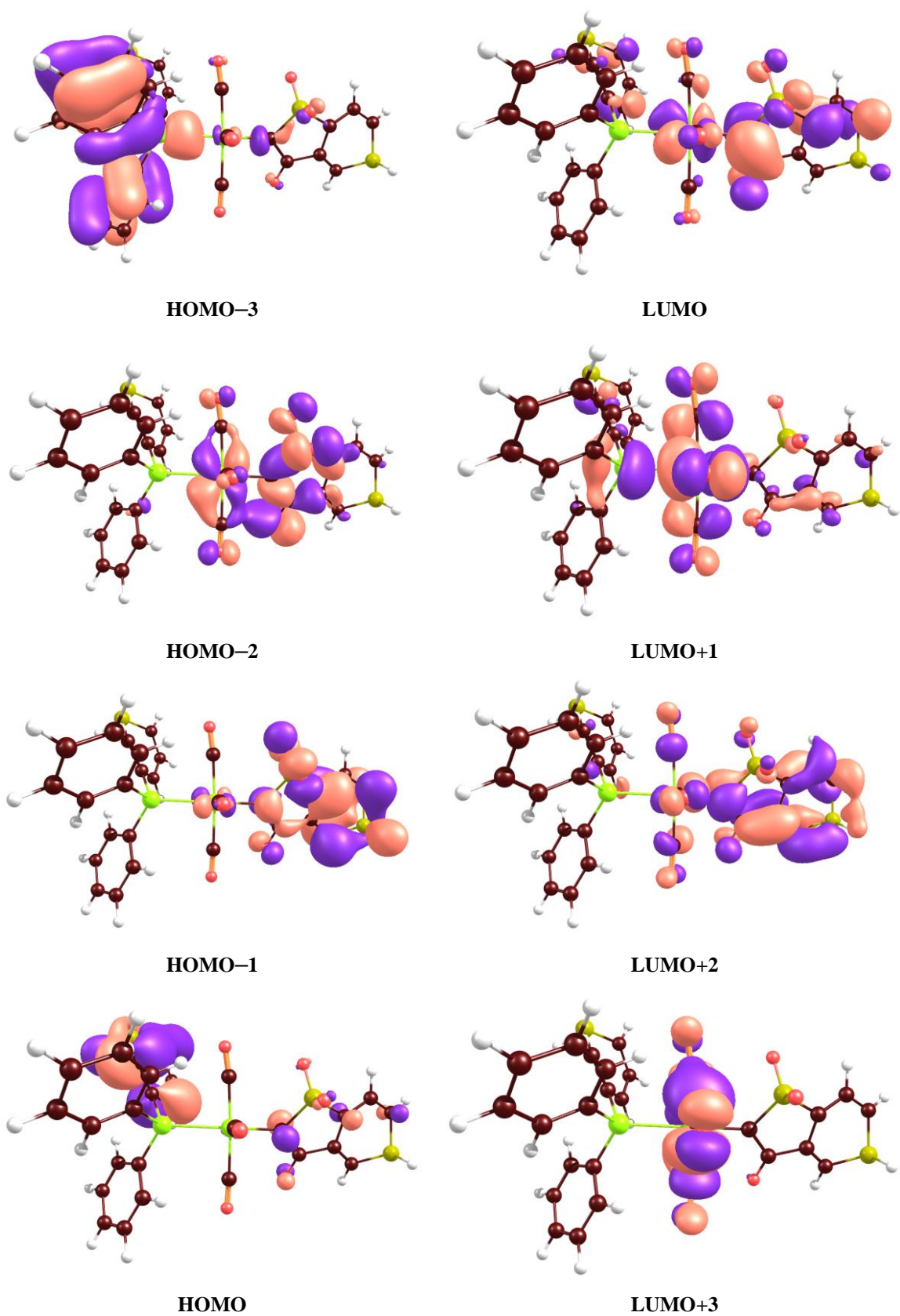

**Fig. 2**. Density functional theory visualized molecular orbitals showcasing the delocalized electron density over M1$^+$ molecule.

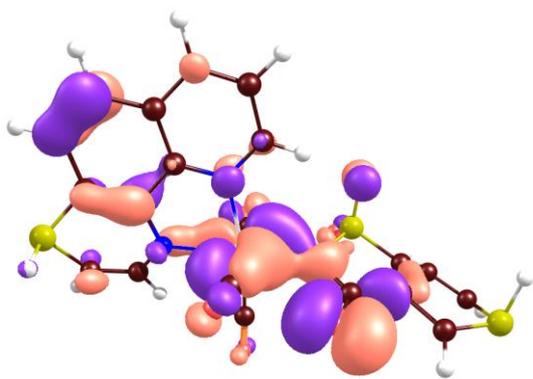

**HOMO–3**

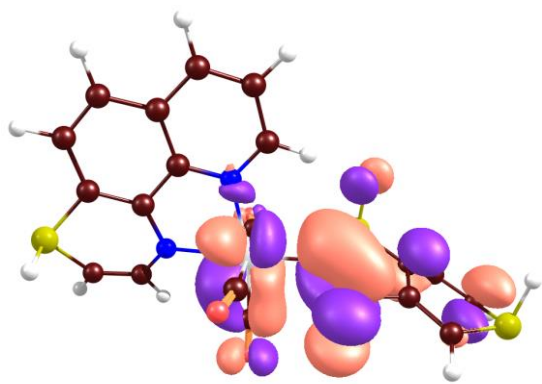

**LUMO**

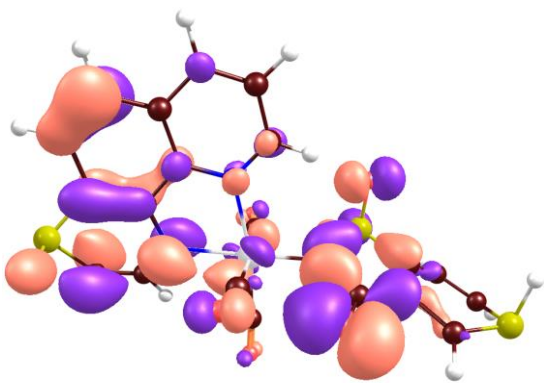

**HOMO–2**

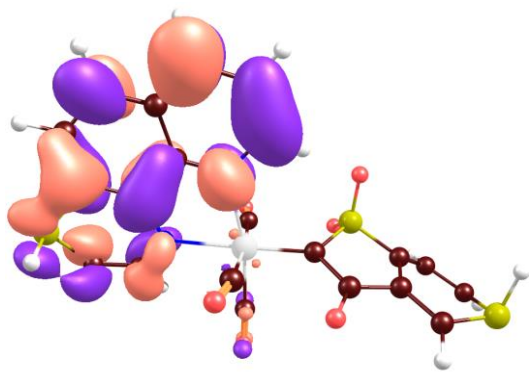

**LUMO+1**

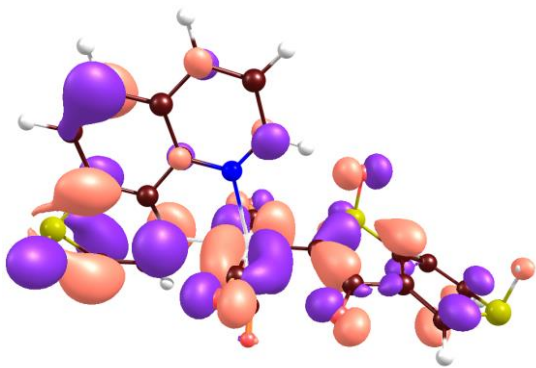

**HOMO–1**

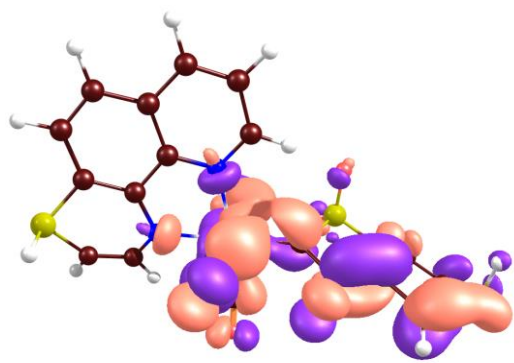

**LUMO+2**

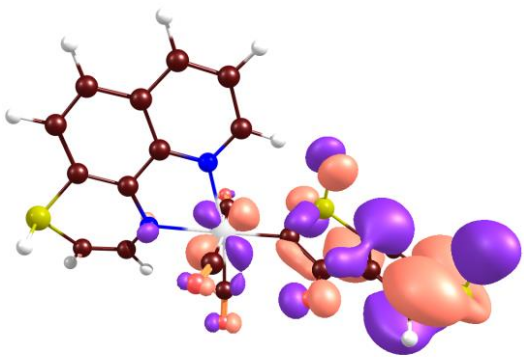

**HOMO**

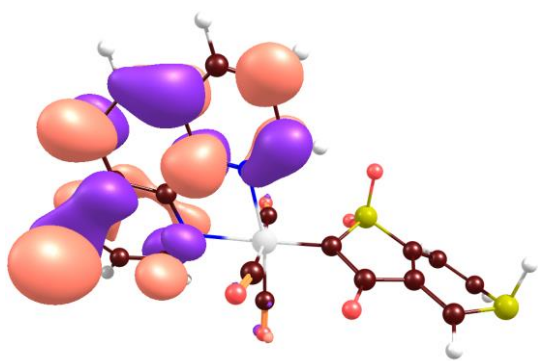

**LUMO+3**

**Fig. 3**. Density functional theory visualized molecular orbitals showcasing the delocalized electron density over M2$^+$ molecule.

We calculated reorganization energies ($\lambda_{o/r}$ and $\lambda_{r/o}$) for M1 and M2 molecules as follows:

$$\lambda_{o/r} = energy(0, +) - energy(0,0); \qquad (i)$$

$$\lambda_{r/o} = energy(+,0) - energy(+,+). \qquad (ii)$$

Here energy (0, +) is the single-point energy of M$^0$ molecule calculated at geometry optimized for M$^+$; energy (0,0) is the optimized energy of M$^0$; energy (+,0) is the single-point energy of M$^+$ molecule calculated at geometry optimized for M$^0$; energy (+,+) is the optimized energy of M$^+$. The results are collected in **Table 3**.

**Table 3.** Reorganization energies (eV) were calculated for M1 and M2 molecules with formulas (i) and (ii).

| Energy | M1 | M2 |
|---|---|---|
| $\lambda_{o/r}$, eV | 0.19 | 0.26 |
| $\lambda_{r/o}$, eV | 0.17 | 0.21 |

**Molecule-electrode interfacial coupling**

To estimate the molecule-electrode electronic coupling, we consider extended molecules M1-Au and M2-Au, containing two Au$_{19}$ atoms, attached to sulfur atoms via S-Au covalent bonds.[54] They are presented in **Fig. 4.** For each $i$-th molecular orbital MO$_i$ ($i$ = 1 to 8), we calculated its coupling strength between the molecule and the left gold electrode, $\Gamma_L$ with as[43,55]

$$\Gamma_L = \pi |M_i|^2 DOS(\varepsilon_i). \qquad (iii)$$

The $M_i$ value is estimated as

$$M_i \approx \varepsilon_i < MO_i | \psi_{golg} >. \qquad (iv)$$

Here, DOS($\varepsilon_i$) is the electronic density of states in bulk gold, separately calculated and presented in **Fig. 5**; <MO$_i$|Ψ> is the overlap between the considered molecular orbital and corresponding HOMO orbital of the left golden nanoparticle included in the extended molecule. The coupling strength between the molecule and the right gold electrode, $\Gamma_R$ was calculated similarly. The results are presented in **Table 4**.

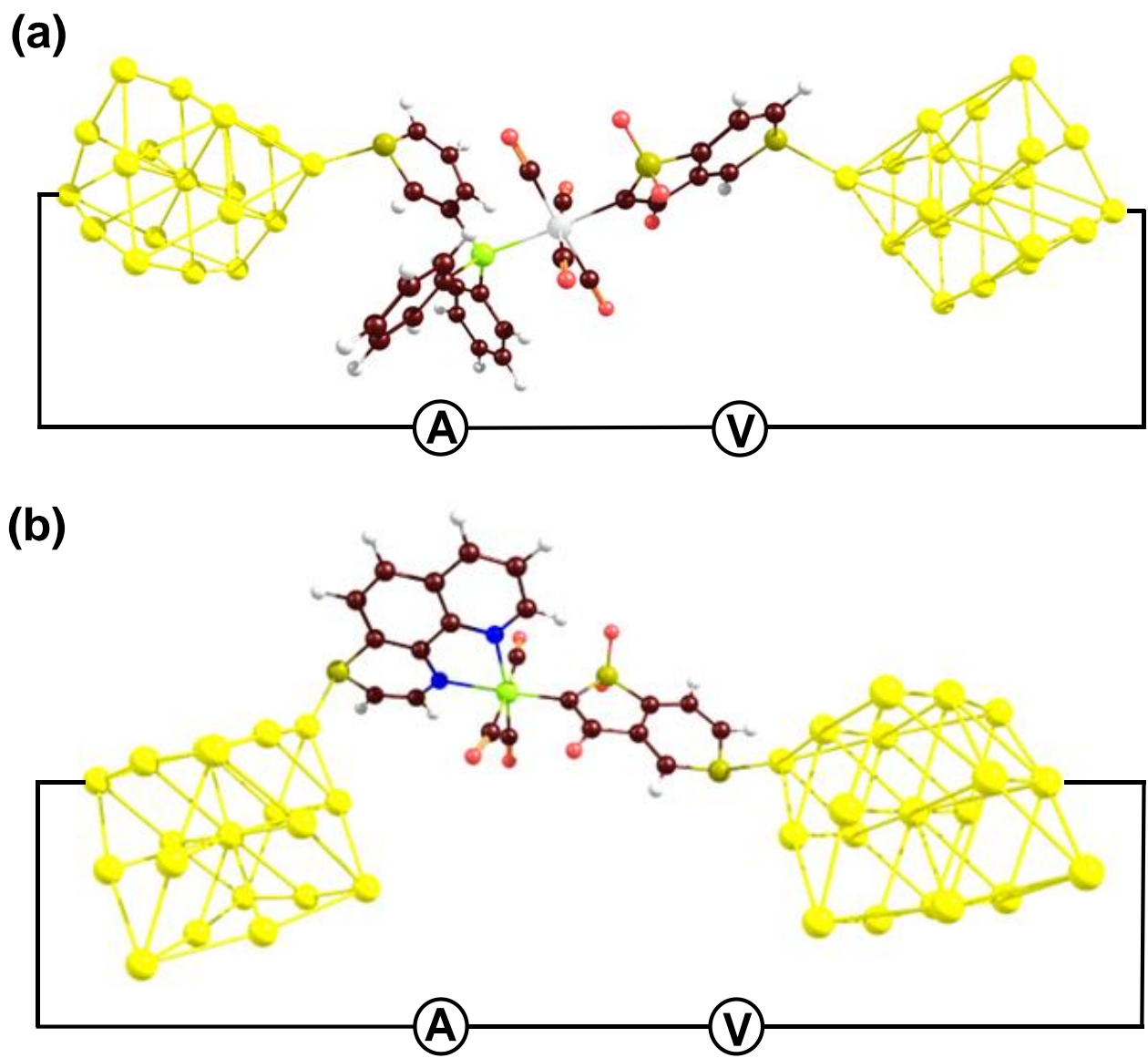

**Fig. 4.** Depiction of single-molecule junctions (a) Au/M1/Au, and (b) Au/M2/Au containing $Au_{19}$ gold atoms as left and right electrical contacts.

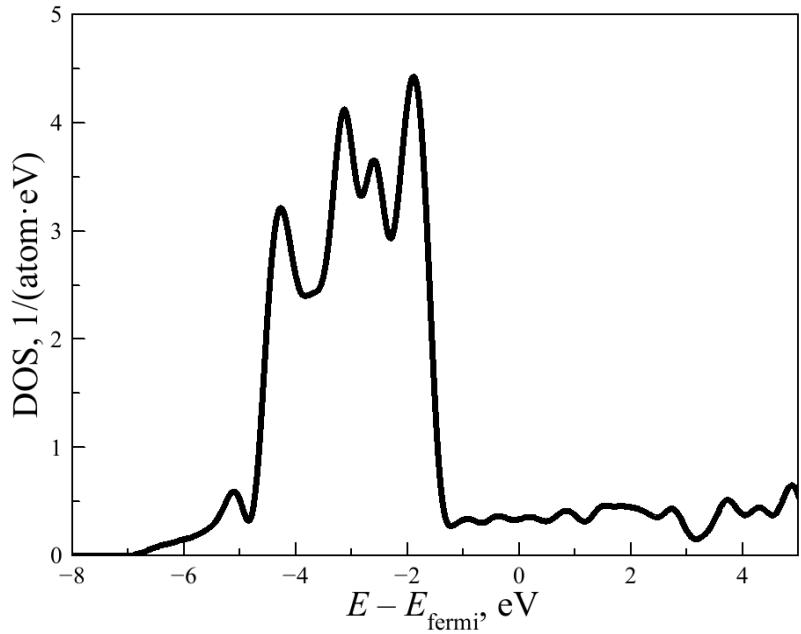

**Fig. 5**. Calculated electronic density of states (DOS) of bulk fcc gold crystal.

**Table 4**. Coupling strength of the molecular orbitals of M1 and M2 molecules with gold electrodes, calculated with formula (iii).

| $i$, # of the orbital | 1 | 2 | 3 | 4 | 5 | 6 | 7 | 8 |
|---|---|---|---|---|---|---|---|---|
| **M1 molecule** | | | | | | | | |
| $\varepsilon_i$, eV | −8.65 | −8.38 | −7.73 | −7.37 | −7.13 | −5.18 | −5.12 | −5.00 |
| DOS($\varepsilon_i$), 1/(atom·eV) | 4.12 | 3.33 | 3.13 | 4.39 | 2.49 | 0.33 | 0.32 | 0.32 |
| <MO$_i$|Ψ> (left) | 0.0312 | 0.0250 | 0.0271 | 0.0216 | 0.0047 | 0.0134 | 0.0043 | 0.0118 |
| <MO$_i$|Ψ> (right) | 0.0012 | 0.0032 | 0.0026 | 0.0026 | 0.0284 | 0.0098 | 0.0271 | 0.0107 |
| $\Gamma_L$, eV | 0.9423 | 0.4589 | 0.4313 | 0.3503 | 0.0088 | 0.0050 | 0.0005 | 0.0035 |
| $\Gamma_R$, eV | 0.0014 | 0.0075 | 0.0040 | 0.0051 | 0.3206 | 0.0027 | 0.0193 | 0.0029 |
| **M2 molecule** | | | | | | | | |
| $\varepsilon_i$, eV | -8.91 | -8.48 | -7.93 | -7.65 | -6.83 | -5.97 | -5.25 | -5.05 |
| DOS($\varepsilon_i$), 1/(atom·eV) | 3.03 | 3.59 | 3.13 | 3.53 | 0.32 | 0.36 | 0.35 | 0.31 |
| <MO$_i$|Ψ> (left) | 0.0051 | 0.0087 | 0.0104 | 0.0038 | 0.0019 | 0.0218 | 0.0274 | 0.0301 |
| <MO$_i$|Ψ> (right) | 0.0034 | 0.0052 | 0.0067 | 0.0295 | 0.0103 | 0.0021 | 0.0026 | 0.0042 |
| $\Gamma_L$, eV | 0.0196 | 0.0614 | 0.0668 | 0.0094 | 0.0002 | 0.0191 | 0.0227 | 0.0225 |
| $\Gamma_R$, eV | 0.0087 | 0.0219 | 0.0277 | 0.5645 | 0.0050 | 0.0002 | 0.0002 | 0.0004 |

**Voltage division factor**

We assume that an external bias, $V$ was applied to the electrodes that bear the single molecule. Potentials at left lead, Re centre, and right lead are 0, $\alpha V$, and $V$, respectively. We define the voltage division factor $\alpha$ as[43]

$$\alpha = \frac{\text{distance(left Au,Re)}}{\text{distance(left Au,right Au)}}. \quad (v)$$

Here distance (left Au, Re) is the distance between the Re atom and the closest Au atom belonging to the left contact; distance (left Au, right Au) is the distance between the two closest Au atoms, belonging to the left and right contacts, respectively.

**Table 5**. Distances between Re and electrodes (Å) and voltage division factor $\alpha$ were calculated for the M1 and M2 molecules.

| Molecule | M1 | M2 |
|---|---|---|
| Distance (left Au, Re) | 7.866 | 6.257 |
| Distance (left Au, right Au) | 8.842 | 8.785 |
| $\alpha$ | 0.471 | 0.416 |

**Tunneling current in single-molecule junction**

Tunneling current, $I_t$ is estimated as a sum of contributions from considered molecular orbitals $MO_i$. The contribution of each orbital is calculated with the Landauer formula:

$$I_t(V) = \frac{2e}{h} \int \left(f_L(E,0) - f_R(E,V)\right) Tr(E,V) dE \quad (vi)$$

The Fermi-Dirac distribution functions in left and right leads are given as:

$$f_L(E,0) = 1/\left(1 + \exp\left(\frac{E-E_F}{kT}\right)\right) \quad (vii)$$

$$f_R(E,V) = 1/\left(1 + \exp\left(\frac{E-E_F+|e|V}{kT}\right)\right) \quad (viii)$$

Here $E_F$ = –5.53 eV is the Fermi level of gold, $kT$ is the Boltzmann constant multiplied by temperature $T$ (we adopted $T = 300$ K). We assume that the left contact is unbiased, whereas the right contact is biased by $V$. The transmission probability for each orbital was estimated as

$$Tr(E,V) = \frac{\Gamma_L \Gamma_R}{(E-\varepsilon+\alpha|e|V)^2+(\Gamma_L+\Gamma_R)^2/4}. \quad (ix)$$

The contributions of different molecular orbitals and total tunneling current through the considered molecules are shown in **Fig. 6**. Both devices showed symmetrical tunneling current within the applied bias ranges.

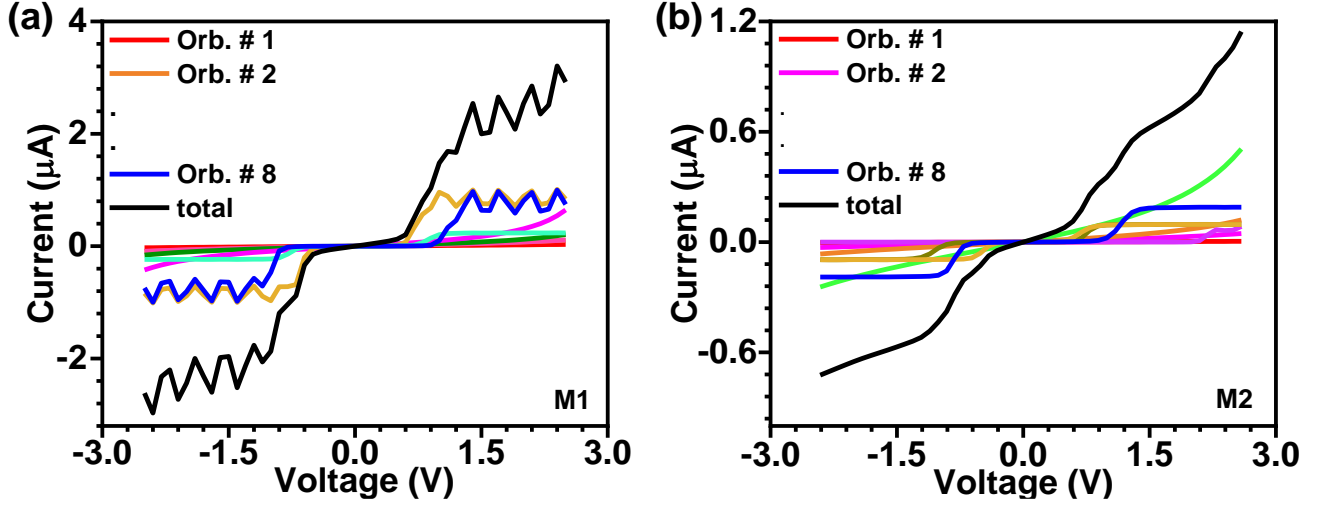

**Fig.6.** Contribution of different molecular orbitals of (a) M1, and (b) M2 molecules containing devices to tunneling current determined based on a single-level model from Landauer-Buttiker formalism.

**Hopping current in the single molecule junction**

The hopping current $I_h$ is associated with the electron hopping between leads and molecules. As a result, the Re(I) compound changes its state from oxidative to redox and vice versa. The current can be easily derived from the hopping rates $R$[42]

$$I_h(V) = -e \frac{R^L_{o/r} R^R_{r/o} - R^L_{r/o} R^R_{o/r}}{R^L_{o/r} + R^R_{r/o} + R^L_{r/o} + R^R_{o/r}}. \qquad (x)$$

Corresponding rates are calculated with the Marcus theory[42,43]

$$R^L_{o/r} = \frac{2\Gamma_L}{h} \sqrt{\frac{\pi}{\lambda_{o/r} kT}} \int f_L(E, 0) \exp\left(-\frac{(E - HOMO(M^0) + \alpha|e|V - \lambda_{o/r})^2}{4\lambda_{o/r} kT}\right) dE \qquad (xi)$$

$$R^L_{r/o} = \frac{2\Gamma_L}{h} \sqrt{\frac{\pi}{\lambda_{r/o} kT}} \int (1 - f_L(E, 0)) \exp\left(-\frac{(-E + HOMO(M^0) - \alpha|e|V - \lambda_{r/o})^2}{4\lambda_{r/o} kT}\right) dE \qquad (xii)$$

$$R^R_{o/r} = \frac{2\Gamma_R}{h} \sqrt{\frac{\pi}{\lambda_{o/r} kT}} \int f_R(E, V) \exp\left(-\frac{(E - HOMO(M^0) + \alpha|e|V - \lambda_{o/r})^2}{4\lambda_{o/r} kT}\right) dE \qquad (xiii)$$

$$R^R_{r/o} = \frac{2\Gamma_R}{h} \sqrt{\frac{\pi}{\lambda_{r/o} kT}} \int (1 - f_R(E, V)) \exp\left(-\frac{(-E + HOMO(M^0) - \alpha|e|V - \lambda_{r/o})^2}{4\lambda_{r/o} kT}\right) dE \qquad (xiv)$$

Here we assumed that during the hopping the electron energy changes from arbitrary value $E$ to the HOMO level of the redox complex $HOMO(M^0) - \alpha|e|V$. The hopping currents through considered molecules calculated by formula (x) are presented in **Fig. 7**.

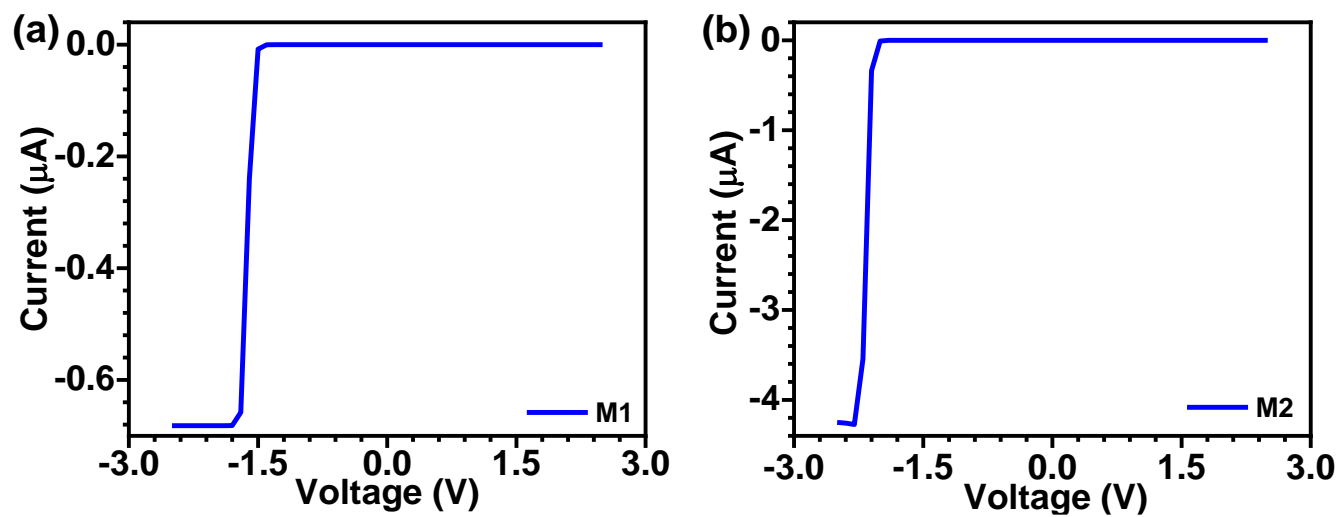

**Fig.7.** Hopping current through the molecules (a) M1, and (b) M2 determined utilizing a single-level model based on Marcus theory.

**Current-voltage features**

The total current $I$ was calculated as a simple sum of the tunneling current $I_t$ and hopping current $I_h$:

$$I = I_t + I_h. \qquad (xv)$$

The resulting I-V curves for both molecules are shown in **Fig. 8.** The curve illustrates that the tunneling current contribution is quite symmetric, while the hopping current contribution provides rectifying characteristics. To obtain stronger rectifying properties, the tunneling contribution should be reduced. This can be achieved, for example, by forming a molecular layer of thickness more than 5 nm in length. Tunneling through such a layer is negligible, so the current is determined by the hopping mechanism only.

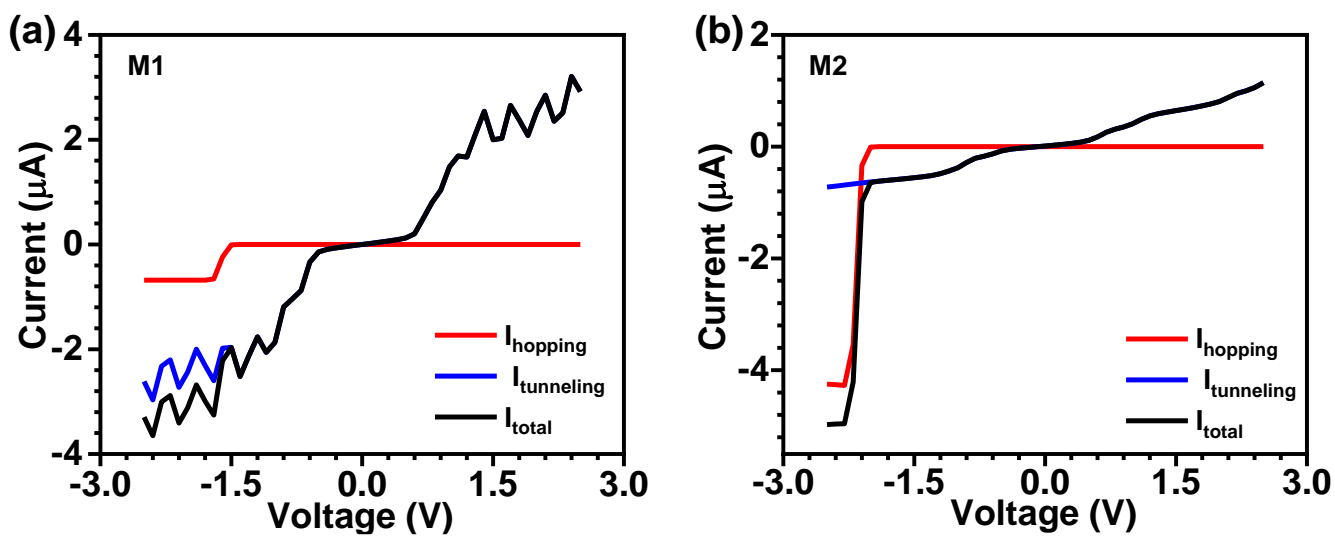

**Fig. 8**. Current-voltage characteristics for the molecules (a) M1, and (b) M2. Contributions of hopping (red line) and tunneling (blue line) to the total current (black line) are presented.

**Conclusion**

In the present work, two Re(I)-carbonyl complexes having different chemical and electronic structures have been theoretically modeled for charge transport properties in Au/Re-complex/Au single-molecule junctions. In redox-active Re(I)-carbonyl complexes both tunneling and hopping mechanisms contribute to the charge transport. However, tunneling is the dominant mechanism for charge transport in Au/M1/Au molecular junctions, while the hopping mechanism dominates in Au/M2/Au molecular junctions also supported by the reorganization energy values calculated as per Marcus theory. Because of different electronic structures, there is variation in the dominant mechanism of charge transport and the nature of current-voltage characteristics. The asymmetry observed in the I-V characteristics of Au/M2/Au molecular junctions is attributed to the dominant hopping mechanism. Landauer−Buttiker formalism and Marcus theory are simple and cost-effective methods for theoretically investigating molecular junctions. Theoretical model studies for less explored Re(I)-complexes can be a useful guide in the experimental design of molecular electronics for understanding current-voltage features and underlying transport models.

**Acknowledgments**

R.K. thanks the University Grant Commission (UGC, New Delhi) for the junior research fellowship. P.C.M. acknowledges the Ministry of Human Resource Development (MHRD), Government of India for SCHEME FOR TRANSFORMATIONAL AND ADVANCED RESEARCH IN SCIENCES (STARS) (Grant No. 2023-0535) for financial support.

## Author Contributions

R.K.: Writing manuscript: S.K.: Computational studies; K.P.K.: Computational studies; writing the manuscript; P.C.M.: Conceptualization, supervising, writing, and editing the manuscript. The manuscript was written through the contributions of all authors. All authors have approved the final version of the manuscript.

## Data availability

The data supporting this article are included as part of the ESI.

## Notes

The authors declare no competing financial interest.

## References

1  S. X. Liu, A. K. Ismael, A. Al-Jobory and C. J. Lambert, *Acc. Chem. Res.*, 2023, **56**, 322–331.

2  E. Tran, M. Duati, V. Ferri, K. Müllen, M. Zharnikov, G. M. Whitesides and M. A. Rampi, *Adv. Mater.*, 2006, **18**, 1323–1328.

3  R. Gupta, J. A. Fereiro, A. Bayat, A. Pritam, M. Zharnikov and P. C. Mondal, *Nat. Rev. Chem.*, 2023, **7**, 106–122.

4  X. Chen, M. Roemer, L. Yuan, W. Du, D. Thompson, E. Del Barco and C. A. Nijhuis, *Nat. Nanotechnol.*, 2017, **12**, 797–803.

5  E. Leary, C. Roldán-Piñero, R. Rico-Sánchez-Mateos and L. A. Zotti, *J. Mater. Chem. C*, 2024, **12**, 4306–4315.

6  R. K. Parashar, P. Jash, M. Zharnikov, and P. C. Mondal, *Angew. Chem. Int. Ed.* 2024, **63**, e202317413.

7  A. Aviram and M. A. Ratner, *Chem. Phys. Lett.*, 1974, **29**, 277–283.

8  R. M. Metzger, *Chem. Rev.*, 2015, **115**, 5056–5115.

9  C. Joachim, J. K. Gimzewski, R. R. Schlittler and C. Chavy, *Phys. Rev. Lett.*, 1995, **74**, 2102–2105.

10  D. J. Wold and C. D. Frisbie, *J. Am. Chem. Soc.*, 2001, **123**, 5549–5556.

11  K. Liu, X. Wang and F. Wang, *ACS Nano*, 2008, **2**, 2315–2323.

12  Z. Li and E. Borguet, *J. Am. Chem. Soc.*, 2012, **134**, 63–66.

13  C. Vericat, M. E. Vela, G. Corthey, E. Pensa, E. Cortés, M. H. Fonticelli, F. Ibañez, G. E. Benitez, P. Carro and R. C. Salvarezza, *RSC Adv.*, 2014, **4**, 27730–27754.

14  L. Newton, T. Slater, N. Clark and A. Vijayaraghavan, *J. Mater. Chem. C*, 2013, **1**, 376–393.

15  E. C. H. Sykes, B. A. Mantooth, P. Han, Z. J. Donhauser and P. S. Weiss, *J. Am. Chem. Soc.*,

2005, **127**, 7255–7260.

16  Y. Zheng, D. Qi, N. Chandrasekhar, X. Gao, C. Troadec and A. T. S. Wee, *Langmuir*, 2007, **23**, 8336–8342.

17  E. C. H. Sykes, P. Han, S. A. Kandel, K. F. Kelly, G. S. McCarty and P. S. Weiss, *Acc. Chem. Res.*, 2003, **36**, 945–953.

18  T. Morita and S. Lindsay, *J. Am. Chem. Soc.*, 2007, **129**, 7262–7263.

19  P. Sachan and P. C. Mondal, *Analyst*, 2020, **145**, 1563–1582.

20  Y. Wei, L. Li, J. E. Greenwald and L. Venkataraman, *Nano Lett.*, 2023, **23**, 567–572.

21  I. Bâldea, *Adv. Theory Simulations*, 2022, **5**, 1-10.

22  L. Li, C. Nuckolls and L. Venkataraman, *J. Phys. Chem. Lett.*, 2023, **14**, 5141–5147.

23  I. Bâldea, *Adv. Theory Simulations*, 2022, **5**, 1–5.

24  X. Song, B. Han, X. Yu and W. Hu, *Chem. Phys.*, 2020, **528**, 110514.

25  P. C. Mondal, U. M. Tefashe and R. L. McCreery, *J. Am. Chem. Soc.*, 2018, **140**, 7239–7247.

26  A. Daaoub, L. Ornago, D. Vogel, P. Bastante, S. Sangtarash, M. Parmeggiani, J. Kamer, N. Agraït, M. Mayor, H. Van Der Zant and H. Sadeghi, *J. Phys. Chem. Lett.*, 2022, **13**, 9156–9164.

27  R. Gupta, P. Jash, P. Sachan, A. Bayat, V. Singh and P. C. Mondal, *Angew. Chem. Int. Ed.* 2021, **133**, 27110–27127.

28  Z. Xie, I. Bâldea and C. D. Frisbie, *J. Am. Chem. Soc.*, 2019, **141**, 18182–18192.

29  H. Yan, A. J. Bergren, R. McCreery, M. L. Della Rocca, P. Martin, P. Lafarge and J. C. Lacroix, *Proc. Natl. Acad. Sci. U. S. A.*, 2013, **110**, 5326–5330.

30  G. Kastlunger and R. Stadler, *Phys. Rev. B - Condens. Matter Mater. Phys.*, 2015, **91**, 125410.

31  A. Migliore and A. Nitzan, *ACS Nano*, 2011, **5**, 6669–6685.

32  R. Gupta, S. Bhandari, S. Kaya, K. P. Katin and P. C. Mondal, *Nano Lett.*, 2023, **23**, 10998–11005.

33  S. H. Choi, B. S. Kim, C. Daniel Frisbie, *Science*, 2008, **320**, 1482–1486.

34  T. Hines, I. Diez-Perez, J. Hihath, H. Liu, Z. S. Wang, J. Zhao, G. Zhou, K. Müllen and N. Tao, *J. Am. Chem. Soc.*, 2010, **132**, 11658–11664.

35  X. Zhao, C. Huang, M. Gulcur, A. S. Batsanov, M. Baghernejad, W. Hong, M. R. Bryce and T. Wandlowski, *Chem. Mater.*, 2013, **25**, 4340–4347.

36  Q. Lu, K. Liu, H. Zhang, Z. Du, X. Wang and F. Wang, *ACS Nano*, 2009, **3**, 3861–3868.

37  R. Kaur, B. Singh, V. Singh, M. Zharnikov and P. C. *Coord. Chem. Rev.*, 2024, **514**, 215872.

38  P. Sachan and P. C. Mondal, *J. Mater. Chem. C*, 2022, **10**, 14532–14541.

39  N. Pizarro, M. Duque, E. Chamorro, S. Nonell, J. Manzur, J. R. De la Fuente, G. Günther, M. Cepeda-Plaza and A. Vega, *J. Phys. Chem. A*, 2015, **119**, 3929–3935.

40  T. Auvray, A. K. Pal and G. S. Hanan, *Eur. J. Inorg. Chem.*, 2021, **2021**, 2570–2577.

41  A. M. Maroń, A. Szlapa-Kula, M. Matussek, R. Kruszynski, M. Siwy, H. Janeczek, J. Grzelak, S. MaćKowski, E. Schab-Balcerzak and B. MacHura, *Dalt. Trans.*, 2020, **49**, 4441–4453.

42  J. K. Sowa, J. A. Mol and E. M. Gauger, *J. Phys. Chem. C*, 2019, **123**, 4103–4108.

43  X. Song, X. Yu and W. Hu, *J. Phys. Chem. C*, 2020, 124, 24408–24419.

44  I. S. Ufimtsev and T. J. Martinez, *J. Chem. Theory Comput.*, 2009, **5**, 2619–2628.

45  A. V. Titov, I. S. Ufimtsev, N. Luehr and T. J. Martinez, *J. Chem. Theory Comput.*, 2013, **9**, 213–221.

46  K. L. Schuchardt, B. T. Didier, T. Elsethagen, L. Sun, V. Gurumoorthi, J. Chase, J. Li and T. L. Windus, *J. Chem. Inf. Model.*, 2007, **47**, 1045–1052.

47  P. J. Hay and W. R. Wadt, *J. Chem. Phys.*, 1985, **82**, 299–310

48  A. D. Becke, *J. Chem. Phys.*, 1992, **96**, 2155–2160.

49  T. Lecklider, *EE Eval. Eng.*, 2011, **50**, 36–39.

50  T. Lu and F. Chen, *J. Comput. Chem.*, 2012, **33**, 580–592.

51  P. Giannozzi, O. Andreussi, T. Brumme, O. Bunau, M. B. Nardelli, M. Calandra, R. Car, C. Cavazzoni, D. Ceresoli, M. Cococcioni and others, *J. Phys. Condens. Matter*, 2017, **29**, 465901.

52  P. Giannozzi, S. Baroni, N. Bonini, M. Calandra, R. Car, C. Cavazzoni, D. Ceresoli, G. L. Chiarotti, M. Cococcioni, I. Dabo, A. Dal Corso, S. De Gironcoli, S. Fabris, G. Fratesi, R. Gebauer, U. Gerstmann, C. Gougoussis, A. Kokalj, M. Lazzeri, L. Martin-Samos, N. Marzari, F. Mauri, R. Mazzarello, S. Paolini, A. Pasquarello, L. Paulatto, C. Sbraccia, S. Scandolo, G. Sclauzero, A. P. Seitsonen, A. Smogunov, P. Umari and R. M. Wentzcovitch, *J. Phys. Condens. Matter*, 2009, **21**, 395502.

53  A. Dal Corso, *Comput. Mater. Sci.*, 2014, **95**, 337–350.

54  T. Mori and T. Hegmann, *J. Nanoparticle Res.*, 2016, **18**, 1–36.

55  Z. Q. You and C. P. Hsu, *Int. J. Quantum Chem.*, 2014, **114**, 102–115.